\documentclass[]{article}
\usepackage{lineno,hyperref}

 \usepackage{setspace}
\usepackage{tocloft}

\usepackage{amsmath,amssymb,amsfonts}
\usepackage{color}
\usepackage{enumerate}
\usepackage{graphicx}   
\usepackage{epstopdf}
\DeclareGraphicsExtensions{.pdf,.eps,.jpg,.mps}
\usepackage{subcaption}
 
\usepackage{hyperref}
\hypersetup{
 colorlinks,
 citecolor=blue,
 }
\usepackage{url}

\title{Maximal visualization-enhancement of latent
fingermarks on polymer banknotes using columnar thin films}
\author{Muhammad Faryad$^{a,b,\ast}$, Akhlesh Lakhtakia$^a$\\\\
$^a$The Pennsylvania State University, \\Department of Engineering Science and Mechanics, \\University Park, PA 16802-6812,   USA.\\
$^b$Lahore University of Management Sciences, \\Department of Physics, Lahore 54792, Pakistan.\\
$^\ast$\texttt{muhammad.faryad@lums.edu.pk}}

\begin{document}

\maketitle

\begin{abstract}
Polymer banknotes are being increasingly adopted to replace older  banknotes. Since banknotes are forensically important substrates for fingermark detection and identification, we present a single-step process to enhance the visualization of fingermarks on banknotes using columnar thin films (CTFs) of nickel. This single-step vacuum technique enhances the quality grade of fingermarks maximally, whether the fingermarks are aged for  one or seven days before CTF deposition. This work represents progress over currently available   sequences of diverse techniques  for enhancing fingermarks on polymer banknotes.  
\end{abstract}



\section{Introduction}
Polymer banknotes are replacing older cotton-fiber- and paper-based banknotes worldwide because polymers are more durable, allow the incorporation of modern anti-counterfeiting features, and are better for the environment \cite{IMF}.  Polymer banknotes started appearing in trials during the 1980s in Haiti, the first completely polymer banknotes being issued by Australia in 1988 \cite{IMF, Haiti}. 

Banknotes are   common substrates for finding and detecting fingermarks for obvious reasons.  Many fingermarks are latent, i.e., they cannot be seen clearly enough by fingermark examiners to identify persons of interest. Diverse physical and chemical techniques have been devised to develop latent fingermarks for visualization \cite{Lee}, the choice of the development technique dependent on the substrate because not all techniques work well enough on all forensically relevant substrates. On some substrates, a sequence of two or more techniques may have to be used.

The first preliminary study on the visualization-enhancing development of   latent fingermarks on polymer banknotes was published early in 1999 \cite{Flynn}. This Australian study encompassed several sequences of techniques:  fluorescent-powder dusting \cite{Lee} or cyanoacrylate fuming \cite{Watkin} for $10-12$~h followed by the vacuum-metal-deposition (VMD) technique \cite{Batey} and fluorescent staining. Certain sequences were  found to enhance the visualization of latent fingermarks on Australian banknotes  aged for a few hours to four days or six  months. However, the enhancement of fingermarks using a chosen sequence was dependent upon the position of the fingermark on the banknote and the duration of environmental exposure.

After that Australian study, the major work on polymer banknotes came out in 2014 from Canada \cite{Lam-I,Lam-II}. In this  study, the fingermarks aged for one, seven, or fourteen days were enhanced with various sequences of dvelopment techniques. The best enhancement of fingermarks was obtained  using a sequence of cyanoacrylate fuming, VMD, fluorescent treatment, and staining with a dye. However, the major bottleneck in this work is that   one has to wait for at least a week and as much as two weeks (depending on the fingermark's condition) after cyanoacrylate fuming before the VMD technique is implemented for best enhancement, thereby
 introducing a long delay in  identification.

Later, another preliminary study from the United Kingdom (UK) in 2016 \cite{Davis} showed that a sequence of superglue fuming, VMD, and superglue fluorescent fuming can be used to enhance most of the fingermarks on polymer banknotes issued by the Bank of England. Other studies from the UK have proposed other sequences involving various powders for enhancing the visualization of fingermarks on the polymer banknotes issued by the Bank of England \cite{Downham2017,Downham2018},  Clydesdale Bank, and the Royal Bank of Scotland \cite{Joan}. The UK Home Office guideline on visualizing fingermarks on polymer banknotes is also a sequence of powder treatments followed by gelatin lifting \cite{HomeOffice}.

Since all current development procedures for polymer banknotes involve a sequence of techniques, with at least three techniques in a sequence, the visualization of fingermarks on polymer banknotes is a lengthy process which might even involve toxic powders. Therefore,  we undertook a pilot study \cite{IFRG}   that involves a single-step procedure of depositing a columnar thin film (CTF) on polymer banknotes to get high-quality fingermarks for visualization. The  CTF technique  has already been used  for visualization of latent fingermarks on several other substrates \cite{Shaler2011,Lakhtakia2011}. In this technique,  a CTF comprising upright nanocolumns
of an appropriate material is deposited on top of a rapidly rotating substrate with the fingermark, this development technique being implemented in a vacuum chamber   \cite{Steve2014}, and the apparatus being the same as   used for the VMD technique \cite{Steve2015}.   

For this preliminary study, three types of polymer banknotes were selected and a single donor was used for all fingermarks.
 The details of fingermark acquisition, CTF deposition, photography, and the grading scheme for fingermark quality are presented in Sec. \ref{methods}. The results are presented and discussed in Sec. \ref{results}, followed by concluding remarks   in Sec. \ref{conc}.

\section{Materials and Methods}\label{methods}
We sourced minimally used polymer banknotes from circulation in public for the work reported here. The denominations of the polymer banknotes chosen were \$$5$ of the Bank of Canada, \pounds$10$ of the Bank of England, and RM~$1$ of the Central Bank of Malaysia. The banknotes were cut into smaller pieces of $3\times2$ cm$^2$ size and pasted onto glass slides with double-sided Kapton tape (S-14532, Uline, Pleasant Prairie, WI, USA). 

The sole donor used the same fingertip to place all fingermarks after rubbing that fingertip on their face and behind the ears for $10$~s to collect sebaceous sweat. Both the obverse  and reverse faces of the banknotes were used for  fingermark placement. Four samples were prepared from each banknote, two from the obverse face and two from the reverse face. Twelve different samples were thus acquired.
One of two samples from each face was aged for one day and the other for seven days in typical spring office environment (temperature $\sim$ 21~$^\circ$C, relative humidity $\sim$ 40-50\%).

CTFs of nickel were deposited using thermal evaporation \cite{Mattox} in a vacuum chamber (Torr international, New Windsor, NY, USA) with a base pressure of $85\pm15$~$\mu{\rm Torr}$. A  tungsten boat (S4-.005W, R. D. Mathis, Long Beach, CA, USA) containing a piece of nickel wire (EVMNI38020, K. J. Lesker,
Jefferson Hills, PA, USA) was heated by passing an electric current of $75$ A through the boat  to generate a  collimated vapor flux of nickel.  This  vapor flux was directed towards the sample mounted on a planar platform
 that was tilted at an angle of  $20$~deg with respect to the arriving vapor flux.  The platform was affixed to a substrate holder. During the deposition, the substrate holder was rotated at the rate of $180$~rpm about an axis passing normally through the center of the substrate holder, while a quartz crystal monitor (SPC-1093-G10, Inficon, Syracuse, NY, USA) was used to maintain the CTF deposition rate  in a narrow range around $0.05$ nm~s$^{-1}$.  The substrate tilt angle of $20$~deg and the rotation rate of $180$ rpm were chosen based on  earlier quality-grade optimization studies on several different forensically relevant substrates \cite{Muhlberger2014,Nagachar2020}. 
Deposition was stopped after the quartz crystal monitor indicated that a CTF of the chosen thickness had been deposited.
The sample was left to cool in the chamber for about 30~min  after the CTF deposition before venting the chamber. The total time from placing the sample into the chamber, pumping the chamber down to the required pressure, deposition, cooling, and venting the chamber to get the samples out ranged from $3$ to $4$ h.

The CTF thickness  of $25$ nm was chosen after a quality-grade optimization study conducted for the polymer banknotes over the $7-40$~nm range, as measured using the DEKTAK 6M stylus profilometer (Bruker, Billerica, MA, USA). 

Immediately before and immediately after CTF deposition, every sample was photographed by a Nikon~D90 camera with a Nikon  105~mm lens. The mode dial was  set to \textsc{close-up} pictures. All photographs were taken with autofocus turned on and  the sample-to-camera distance   was kept the same for all samples.
The sample was illuminated with a Sylvania 100 W incandescent bulb at an angle that gave the best possible image of the fingermark for each sample. Also, the photographs were taken either perpendicularly or at a small tilt to get the best images. In Sec. \ref{results}, the best images of the fingermarks are presented before and after the CTF deposition. 

Every one of the 12 fingermarks was graded by one of us (M.F.) for visual quality before and after CTF deposition  with a grading scheme used extensively by the UK Home Office \cite{Sears}. According to this scheme, the fingermark is given an integer grade from $0$ to $4$. A grade of $0$ indicates no evidence of a fingermark whereas a grade of $4$ indicates  complete ridge detail of the whole fingermark. A grade of $1$ indicates  evidence of contact of the substrate with the finger but no visual evidence of the ridges. A grade of $2$ indicates that ridges are visible in about a third of the fingermark, whereas a grade of $3$ indicates the ridge visibility for about two thirds of the fingermark. Only a fingermark with a grade of $3$ or $4$ can be used to reliably identify the person.

\section{Results and Discussion}\label{results}
\begin{figure}[htbp]
\centering
\includegraphics[width=1.0\linewidth]{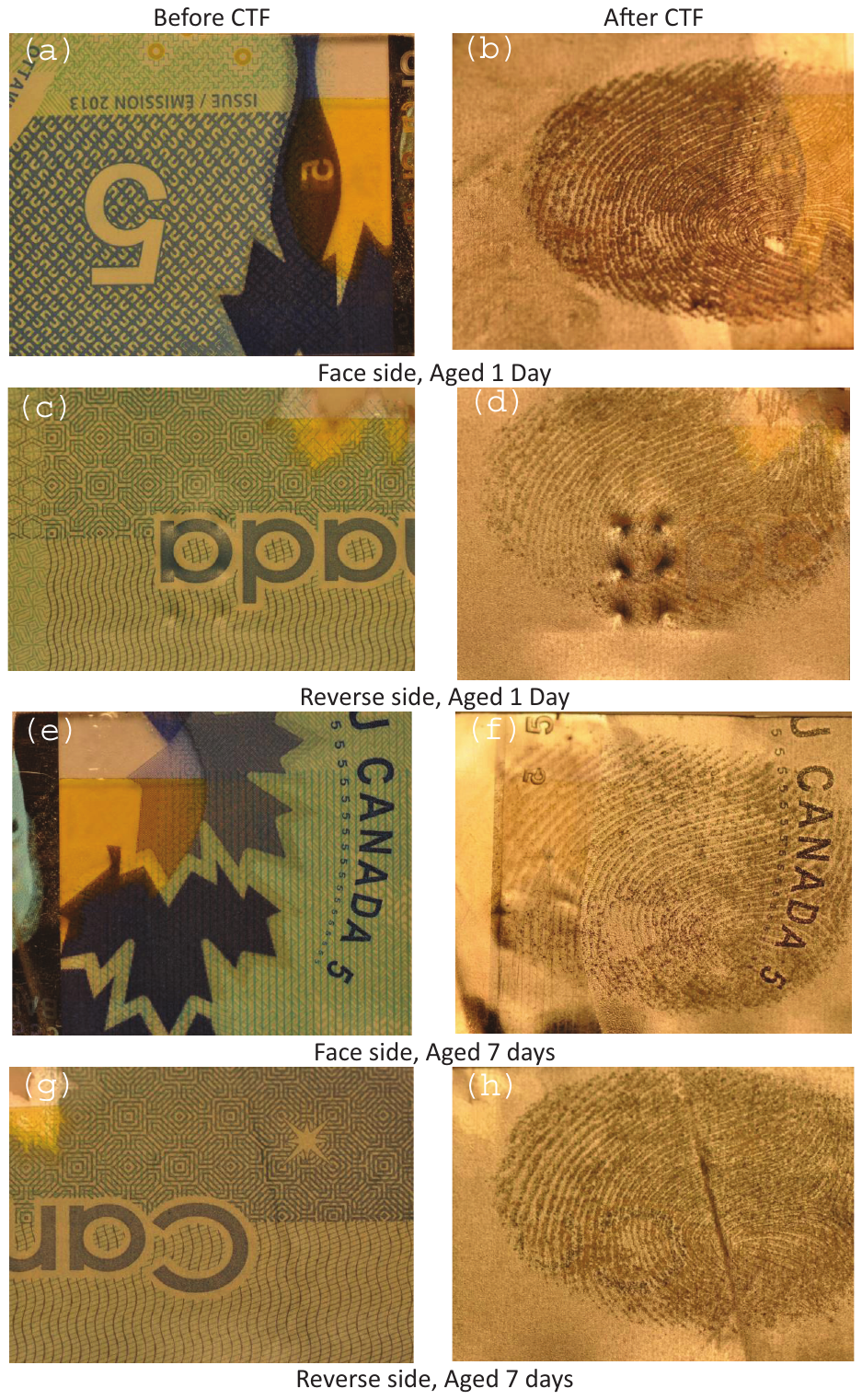}
\caption{ Photographs of latent fingermarks on the obverse and   reverse faces of a \$$5$ polymer banknote (issued by the Bank of Canada)  before and after the deposition of $25$-nm-thick nickel CTF. The fingermarks were aged for either one or seven days before CTF deposition. For all four samples, the quality grade  is $0$ before  and $4$ after the deposition of the CTF.}
\label{fig1}
\end{figure}

Photographs of four samples aged for either one day or seven days before CTF deposition and developed with a $25$-nm-thick CTF of nickel are shown in Fig. \ref{fig1} for a \$$5$ polymer banknote issued by the Bank of Canada. The fingermarks were placed on both faces of the banknote. The figure shows that the quality grade of fingermarks before CTF deposition for all four samples is $0$ as the examiner cannot discern if a fingermark is present or not. However, after the single-step CTF deposition, the fingermark is visible completely with enough detail to identify the pattern of ridges in it. Therefore, the quality grade of all samples after CTF deposition is $4$. This maximal enhancement in fingermark quality is due to the reproduction of ridge-and-valley topography of the fingermark by the CTF as well as due to masking by the thin metallic film of the visual features printed but not embossed on the banknote.

In Figs. \ref{fig1}(a)-(g), the yellow color seen in some parts of the banknote is because of the Kapton tape showing from the transparent parts of the banknote. The 3$\times$2 rectangular array of prominent dots seen in Fig. \ref{fig1}(d) is due to the embossed dots in the original banknote intended to guide visually impaired persons. Also, a linear crease present in the banknote before CTF deposition in Fig. \ref{fig1}(g) is exaggerated in Fig. \ref{fig1}(h). The exaggeration of the dots and creases is due to the deposition of the warm-nickel CTF  over the polymer banknote. Let us also note that the colorful patterns of the banknotes get hidden by the nickel CTF except those prints that are embossed into the banknotes, as in Fig. \ref{fig1}(f), because nickel being a metal reflects visible light and does not let it pass through to the colorful face of the banknote. Thus, the CTF technique  enhances the visualization of  fingermarks also by eliminating the effect of colorful patterns on  banknotes.

\begin{figure}[htbp]
\centering
\includegraphics[width=1.0\linewidth]{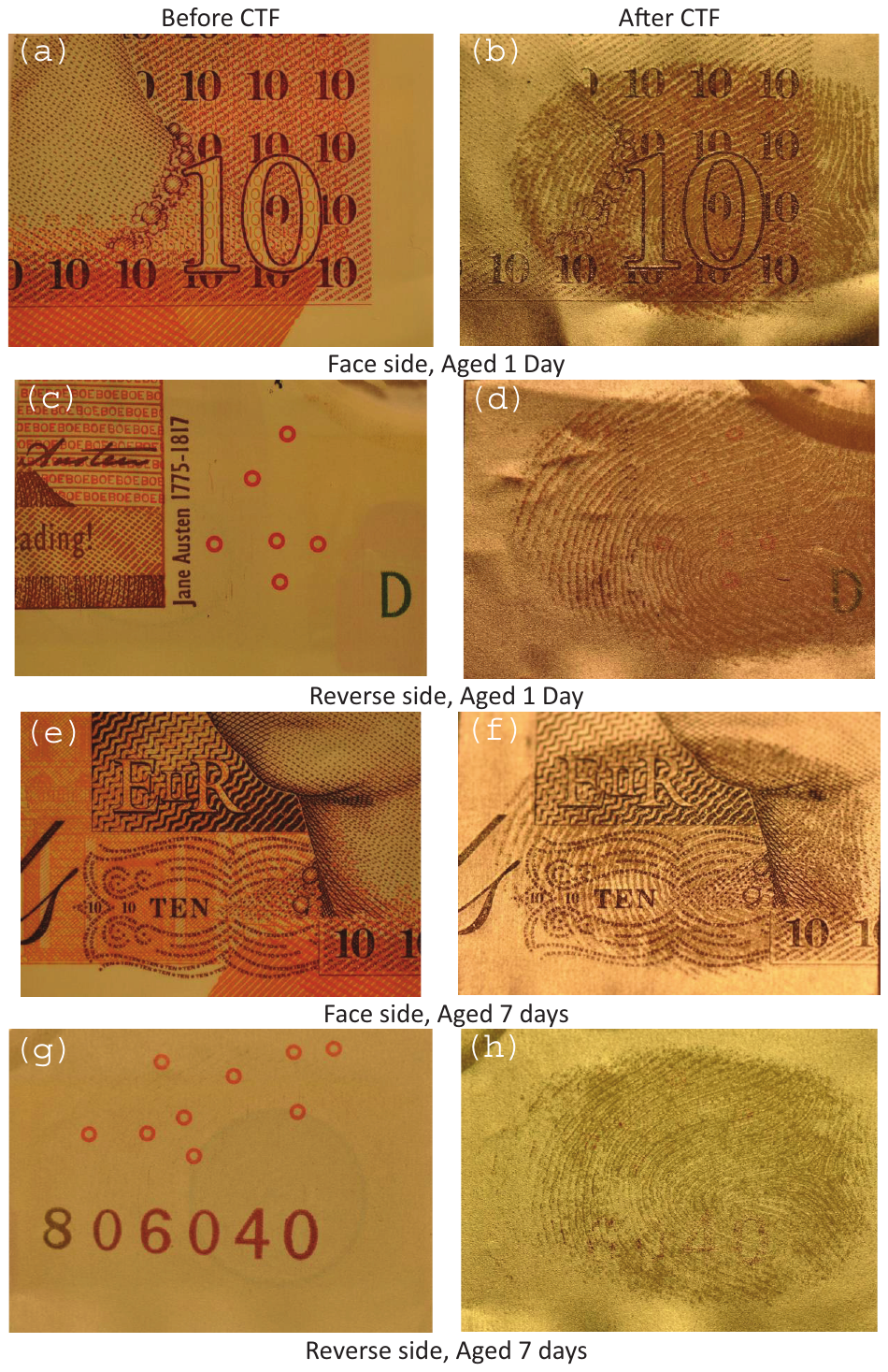}
\caption{ Same as Fig. \ref{fig1} except for a \pounds$10$ polymer banknote issued by the Bank of England.}
\label{fig2}
\end{figure}

\begin{figure}[htbp]
\centering
\includegraphics[width=1.0\linewidth]{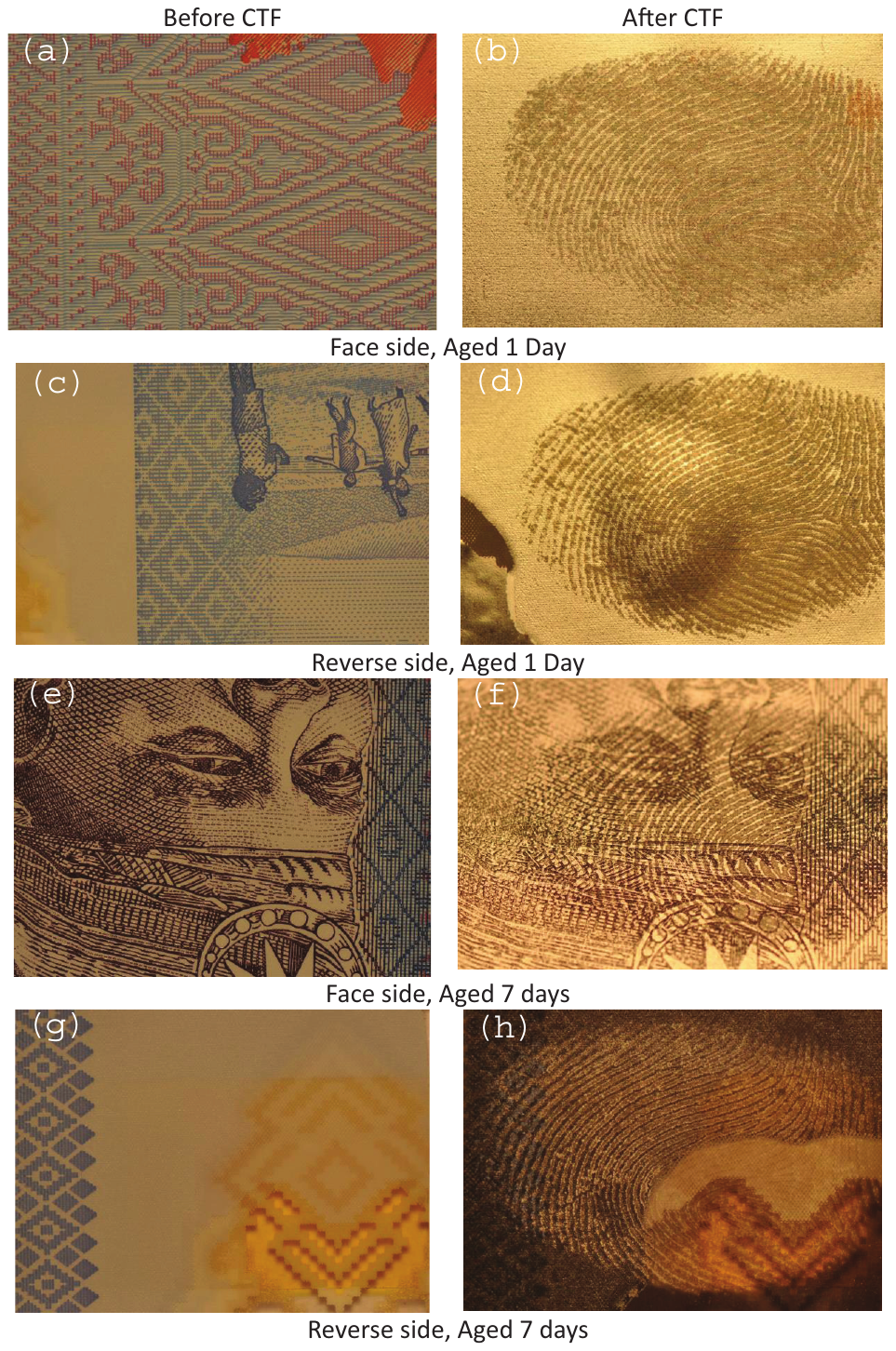}
\caption{Same as Fig. \ref{fig1} except for  a RM~$1$ polymer banknote issued by the Central Bank of Malaysia. }
\label{fig3}
\end{figure}

Figure 2 shows photographs of  fingermarks on a \pounds$10$ banknote issued by the Bank of England. The quality grade of all four samples  before  CTF deposition is $0$ but it improves to $4$ after   CTF deposition. The ability of the nickel CTF to enhance the visual quality of fingermarks but suppress the printed but not embossed features on the banknote is due to the masking of colorful patterns by the thin metallic film and enhancement of the fingermark topography. Since the \pounds$10$ English banknote has considerably more elaborate and rougher patterns than the \$$5$ Canadian banknote, the  fingermarks are slightly less enhanced with respect to the background on the former than on the latter. However, still enough ridge detail is available in the developed fingermarks to given them a grade $4$.

Photographs of four samples of the RM~$1$ Malaysian banknote are given in Fig. \ref{fig3}. Again, CTF deposition enhances the quality grade of the fingermarks from $0$ to $4$. In Fig. \ref{fig3}(f), the fingermark visualization is still significantly enhanced despite  it being located partially atop a complex embossed feature. In Fig. \ref{fig3}(h),  the enhancement is less on the transparent part  than on the other parts, but even there the ridge-and-valley topography of the fingermark can be easily discerned. 

A scan of Figs. \ref{fig1}-\ref{fig3} shows that seven-days-aged fingermarks are slightly less enhanced than one-day-aged fingermarks, but the difference is very small as every fingermark is being enhanced to the maximum possible quality grade of $4$ after the deposition of a $25$-nm-thick  CTF.

Parenthetically,  polymer banknotes
can shrink and even melt at temperatures exceeding 120~$^\circ$C \cite{BoEnote}. But Figs. \ref{fig1}-\ref{fig3} prove that the polymer banknotes survived CTF deposition very well, so that the sample temperature 
was quite low during CTF deposition. Also, the  roughness is not the same across the full surface of banknotes. However, the CTF technique seems to work well enough even on the rough parts to enhance the quality grade of the fingermarks to $4$.

\section{Concluding Remarks}\label{conc}
 Latent fingermarks  on polymer banknotes issued by the Bank of Canada, the Bank of England, and the Central Bank of Malaysia were developed by depositing  nickel CTFs      in a vacuum chamber. The fingermarks were aged for either one day or seven days and placed on either the obverse or the reverse faces of the banknotes. The CTF was deposited in a single step and no pre- or post-processing of fingermarks was done. Also, the photographs of the samples were not enhanced using image-processing software. 
 
 The quality grade of every fingermark increased from $0$ to $4$ after CTF deposition.
 The maximal enhancement of the quality grade of fingermarks on three different types of polymer banknotes shows the potential of CTF technique for enhancing fingermarks on other polymer banknotes also. Although we used thermal evaporation to deposit CTFs, other vacuum techniques such as sputtering \cite{Mareus,Wang} can also be used.
 Let us note at the end that the pilot study presented in this paper is only the first step and a comprehensive study needs to be designed before deciding to use this technique for forensic evidence in a court of law.
 
 \subsection* {Acknowledgments}
 We thank Chengzhi Li for assistance with the DEKTAK 6M stylus profilometer (Bruker, Billerica, MA, USA) for measuring the thickness of a CTF.
MF thanks Lahore University of Management Science for one year of sabbatical leave and the Department of Engineering Science and Mechanics (Penn State) for hosting him during this period.   AL thanks the Charles Godfrey Binder Endowment at Penn State for partial financial support.  This research was substantially funded by the US Department of Homeland Security under Grant Award No. 17STCIN00001-05-00.
 

\end{document}